\documentclass[prl,reprint,showpacs, superscriptaddress]{revtex4-1}
\usepackage{upgreek}
\usepackage{epstopdf}
\usepackage{graphicx}

\begin{document}
\title{Experimental x-ray ghost imaging}
\author{Daniele Pelliccia}
\email[]{daniele.pelliccia@rmit.edu.au}
\affiliation{School of Science, RMIT University, Victoria 3001, Australia}
\affiliation {Australian Synchrotron, Victoria 3168, Australia}
\affiliation{School of Physics and Astronomy, Monash University, Victoria 3800, Australia}
\author{Alexander Rack}
\affiliation{European Synchrotron Radiation Facility, 38043 Grenoble, France}
\author{Mario Scheel}
\affiliation{Synchrotron Soleil, 91192 Gif-sur-Yvette, France}
\author{Valentina Cantelli}
\affiliation{Helmholtz-Zentrum Dresden-Rossendorf, 01328 Dresden, Germany}
\affiliation{European Synchrotron Radiation Facility, 38043 Grenoble, France}
\author{David M. Paganin}
\affiliation{School of Physics and Astronomy, Monash University, Victoria 3800, Australia}

\begin{abstract}
We report an experimental proof of principle for ghost imaging in the hard x-ray energy range. We used a synchrotron x-ray beam that was split using a thin crystal in Laue diffraction geometry. With an ultra-fast imaging camera, we were able to image x-rays generated by isolated electron bunches. At this time scale, the shot noise of the synchrotron emission process is measurable as speckles, leading to speckle correlation between the two beams. 
The integrated transmitted intensity from a sample located in the first beam was correlated with the spatially resolved intensity measured in the second, empty, beam to retrieve the shadow of the sample. The demonstration of ghost imaging with hard x-rays may open the way to protocols to reduce radiation damage in medical imaging and in non-destructive structural characterization using Free Electron Lasers.  
\end{abstract}

\maketitle 
Ghost imaging, in its basic form, is the technique of indirectly imaging a sample by using the correlation between the intensity recorded at two detectors illuminated by spatially separated correlated beams \cite{erkm10}. A bucket detector measures the total intensity transmitted (or scattered) by a sample, placed in one of the beams. The sample image is then retrieved by correlating the output of the bucket detector with a pixel array detector located in the other beam, namely the one that has not directly interacted with the sample. \newline
Initially demonstrated with entangled photon pairs \cite{pitt95}, ghost imaging was subsequently performed using correlation between classical coherent light beams \cite{benn02}. The protocol was shown to be very robust, leading to experimental studies on ghost imaging using pseudo-thermal light \cite{gatt04, cai05, vale05}, true thermal sources \cite{zhan05}, and eventually computational ghost imaging \cite{shap08}, where a computer-controlled spatial light modulator generates a series of known illuminating fields, altogether removing the need for imaging the empty beam. Of relevance for this paper is also a very recent demonstration of Fourier transform ghost imaging using speckle fields generated with partially coherent synchrotron x-rays \cite{yu16}. \newline
At the heart of thermal ghost imaging is the \textit{speckle} correlation in the intensity fluctuations of the illuminating beam. The speckles can be produced either by near--field diffraction of a coherent beam by a slowly moving diffracting object  \cite{gatt04, cai05, vale05, yu16}, or taking advantage of the natural fluctuations of true thermal light \cite{zhan05}, as in the Hanbury Brown--Twiss (intensity) interferometer \cite{hbt}. In this Letter we use the latter mechanism to produce the first proof of principle demonstration of hard x-ray direct ghost imaging using synchrotron emission from an undulator in a third generation synchrotron storage ring. \newline
Synchrotron emission from an ultra-relativistic electron bunch provides a natural thermal source of hard x-rays. Intensity correlation x-ray  experiments, proposed as far back as 1975 \cite{shur75} (see also \cite{ikon92}), were employed several times for coherence characterization of synchrotron \cite{glus92, yaba01,sing14} and x-ray Free Electron Laser (FEL) \cite{sing13} beams. To date though, x-ray speckle correlation has never been used for direct ghost imaging.\\
Such imaging applications are nowadays feasible, given the availability of ultrafast hard x-ray imaging cameras \cite{rack16} that permit spatially resolved measurement in a single frame. By using one such ultrafast detector, coupled to an image intensifier, the light emitted from a single electron bunch is sufficient to form an image containing natural speckles arising from the shot noise of the electron bunch. By splitting the beam into two spatially separated locations on the camera screen, and placing an object in one of the beams, the ghost image of the object can be recovered by a suitable intensity correlation between the two speckle beams.
Demonstrating ghost imaging with hard x-rays is significant, mainly due to a striking peculiarity of the ghost imaging mechanism. Arising from the intensity correlation between separate beams, ghost imaging is remarkably insensitive to turbulence in either beam \cite{meye11}, and applications in atmospheric imaging have followed from this property.  Turbulence is not a problem for hard x-ray imaging, but radiation dose certainly is. The very same idea of robustness to turbulence could be used to make the counting statistics in the two beams very different. In other words the beam that interacts with the sample could be greatly attenuated (with increased associated noise), yet maintaining the intensity correlations with the second, much more intense beam. Therefore, forming ghost images with x-rays that never interacted with a sample is an extremely interesting avenue to mitigate radiation damage. Consequences can be appreciated in medical imaging diagnostics, but also in biological x-ray microscopy where radiation damage often represents the effective limit to the achievable resolution. 
\begin{figure}[t]
\begin{center}
\includegraphics[scale=1.1]{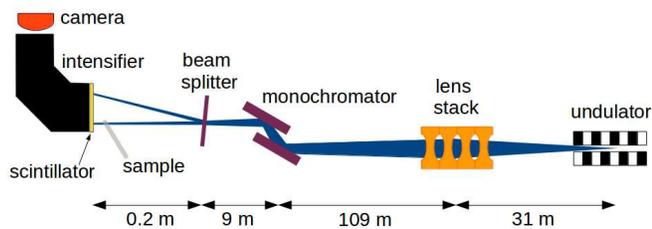}
\end{center}
\caption{Scheme of the experimental setup (not to scale). X-ray beam propagation is from right to left. X-ray pulses from the undulator are focused by a refractive lens stack and then monochromatized by a double-bounce Si monochromator. The beam splitter, working in Laue diffraction, is located at the focal  position of the lenses. Both transmitted and diffracted beams are imaged on the ultrafast camera coupled to a scintillator and an image intensifier.}
\label{fig1}
\end{figure}
Emerging applications of FELs for single molecule diffraction could also benefit from ``diffraction without destruction'' achieved via the ghost imaging mechanism \cite{li15}.  \newline
Our ghost imaging experiment was carried out at the ID19 beamline of the European Synchrotron ESRF in Grenoble (France). We used a special operation mode of the storage ring, in which 
4 equidistant electron bunches are stored, carrying a maximum current of 10 mA/bunch. In this way the temporal separation between the bunches is approximately 704 ns, corresponding to a frequency of about 1.42 MHz. \newline 
The experimental setup is depicted in Fig. \ref{fig1}. The beam from the undulator was focused by a Be refractive lens stack to a focal spot of approximately 1.5 mm $\times$ 1.1 mm (H $\times$ V) at the beam splitter position. The monochromator -- a pair of Si crystals located at 140 m from the source -- monochromatized the beam at an energy of 20 keV. The beam splitter, constituted by a 300 $\upmu$m thick Si crystal polished on both faces, was aligned to select one of the (220) Laue reflections in the forward direction. At the same time, the crystal enabled a portion of the undiffracted beam to be transmitted in the forward direction \cite{bonse1965}. The camera was placed 20 cm downstream of the beam splitter, to be able to detect, within its field of view, both diffracted and transmitted beams. The camera was a Photron FASTCAM SA-Z, coupled to a 200  $\upmu$m thick phosphor screen scintillator (CsI:Na) and an image intensifier \cite{ponchut2001} with a P46 (YAG:Ce) phosphor screen. \newline
Both the refractive lens stack and the image intensifier have been adopted to guarantee sufficient counting statistics to operate the camera at a nominal frame rate $f_c=$ 2.88 MHz, more than twice the storage ring frequency, to ensure correct sampling. In reality, as we discovered during post-processing, the actual frame rate of the camera was lower and equal to 2.57 MHz.   \newline
A typical frame recorded by the Photron camera is shown in Fig. \ref{fig2}(a). The object, a copper wire of 200 $\upmu$m diameter was aligned approximately in the middle of the transmitted beam, which appears in the bottom right corner of Fig. \ref{fig2}(a). The diffracted beam does not contain the object, and is used as a reference beam. A zoom of both the diffracted and transmitted beams is shown in Fig. \ref{fig2}(b) and (c) respectively. In this proof of principle experiment, the intensity of the transmitted beam was integrated over an area of 30 $\times$  20 pixels around the beam centre, to reproduce the 1D (time dependent) bucket signal.
\begin{figure}[t]
\begin{center}
\includegraphics[scale=0.68]{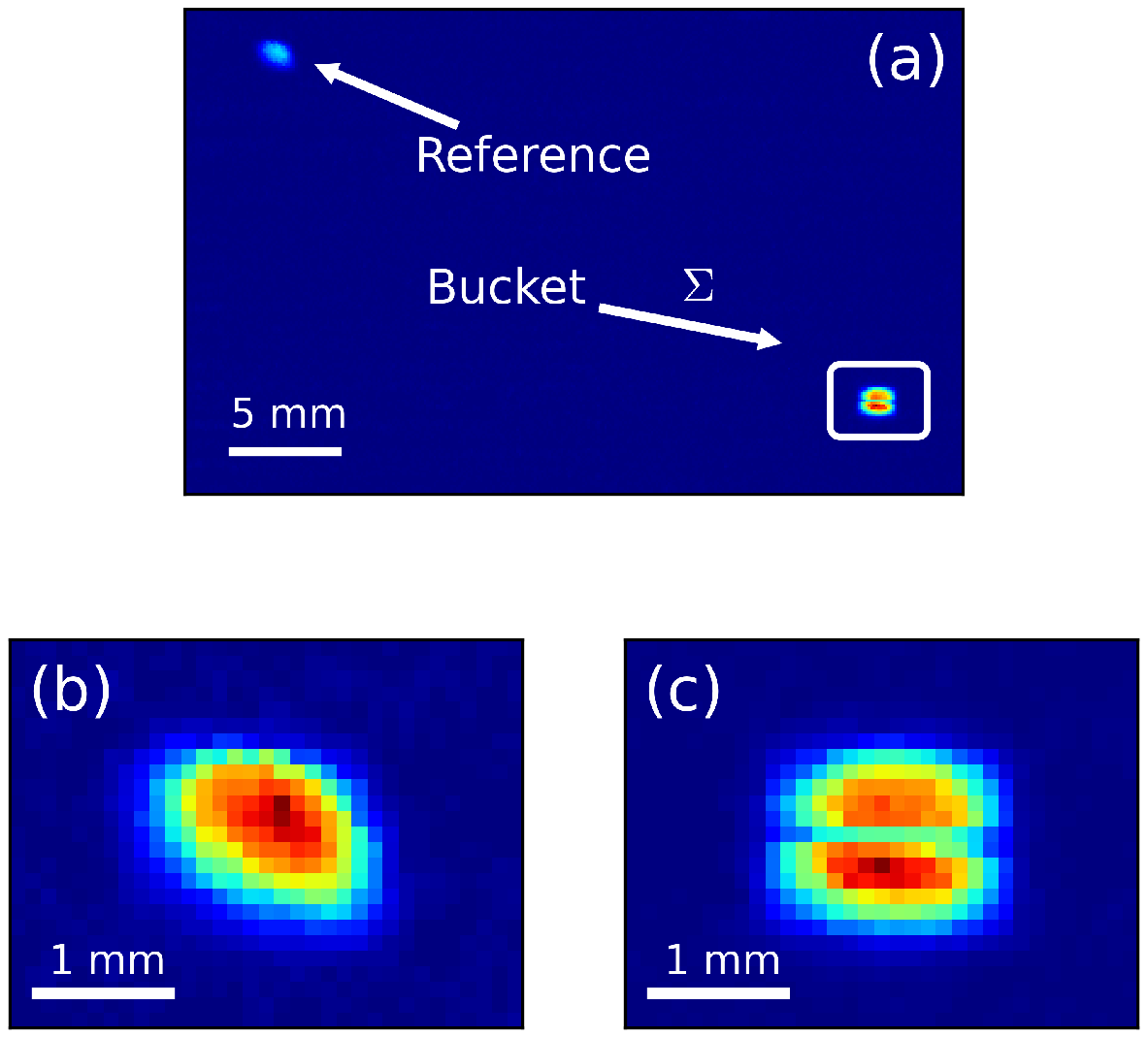}
\end{center}
\caption{Representative frame recorded by the Photron camera. (a) Complete frame with marked position of the reference and bucket beams (diffracted and transmitted beam respectively). The shadow of the wire is visible in the bucket beam. The white frame marks the region that has been integrated over, to obtain the bucket signal $B_{r}$. (b) and (c) show a close-up view of reference and bucket beam (with sample) respectively.} \label{fig2}
\end{figure}
The ghost image $T_{GI}(x,y)$ was then obtained by correlating the bucket signal $B_{r}$ with the reference image $I_{r} (x,y)$ \cite{katz99}:
\begin{equation}
T_{GI}(x,y) = \langle \left( B_{r} - \langle B \rangle  \right) I_{r} (x,y)  \rangle,
\label{eq:gi}
\end{equation}
where $\langle B \rangle$ is the average bucket signal and the averages are calculated over an ensemble of 20000 frames.
\begin{figure}[t]
\begin{center}
\includegraphics[scale=0.65]{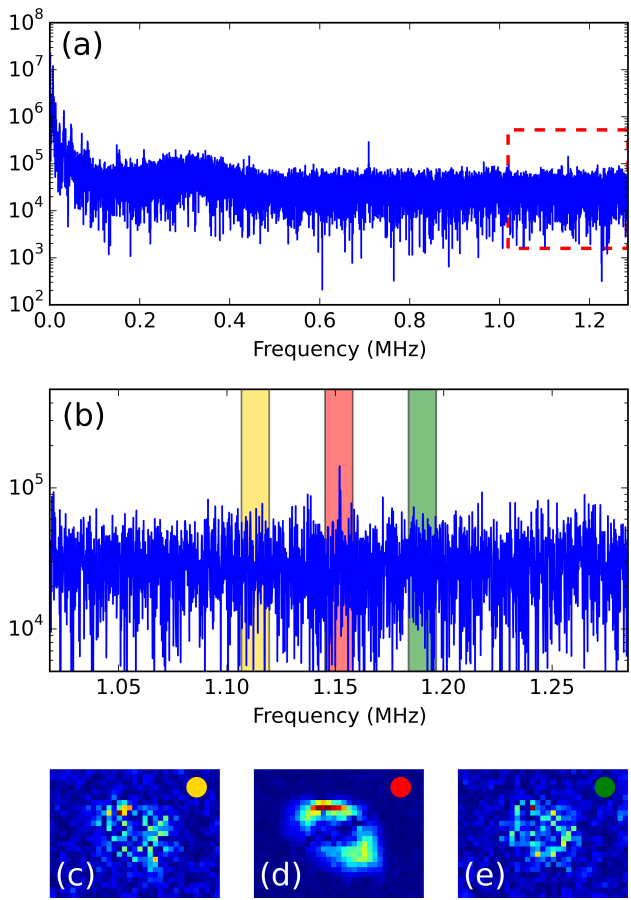}
\end{center}
\caption{The effect of Fourier filtering on the ghost image. (a) Power spectrum of the bucket signal. The low frequency components are related to mechanical instabilities of the crystals. The two sharp peaks visible at higher frequency correspond to half of the ring frequency (0.72 MHz) and the alias of the primary storage ring frequency at 1.15 MHz (see text for details). The region marked by the dashed box is zoomed in in panel (b), where three different windows used for calculation are overlayed. (c-e) Ghost images obtained by Fourier filtering with the windows displayed in panel (b). Ghost imaging is obtained only when the window includes the alias of the storage ring frequency. Only in this case is a true correlation between reference and bucket beam present. The image size in panels (c)-(e) is 3.7 mm $\times$ 2.8 mm (H $\times$ V).} \label{fig3}
\end{figure}
In order to retrieve the ghost image however, Fourier filtering of both bucket and reference signals had to be performed. Both signals are generated in a diffraction process and, due to vibrations of the beam splitter mounting, reference and bucket beam display a low frequency anti-correlation; small changes in the angular position of the beam splitter deviate intensity from the transmitted to the diffracted beam and vice versa. \newline
Such low frequency components, visible in the power spectrum plotted in Fig. \ref{fig3}(a), must be filtered out to isolate the `true' speckle correlation arising from single bunch emission. We originally planned to window the component corresponding to the storage ring frequency $f_{r}$. However, due to the fact that the actual camera frame rate was $f_{c}< 2f_{r}$, the Nyquist frequency $f_{N}=f_{c}/2$ for the system was below the storage ring frequency. As a result the storage ring frequency was not directly accessible, and only an alias at a frequency $ f_{c}-f_{r}=$ 1.15 MHz is visible in the power spectrum. The alias is outlined by the dashed box in Fig. \ref{fig3}(a). In addition, the power spectrum shows a second prominent peak at $f_{r/2}=$ 0.72 MHz, corresponding to half the storage ring frequency. Both peaks can be used for ghost imaging by selecting the frequency components of either the alias or $f_{r/2}$. This corresponds to selecting each x-ray pulse (albeit at down-shifted frequency) or the average of two pulses respectively. In both cases the natural speckle pattern arising from the shot noise of the electron bunches becomes predominant, and therefore produces the true random correlation needed for the ghost imaging. 
\newline
Here we show, as an example, the ghost imaging obtained by windowing the alias of the storage ring frequency at 1.15 MHz. A closeup view of the power spectrum around the alias position is plotted in Fig. \ref{fig3}(b). To demonstrate the use of the speckle correlation, we performed the windowing around three different frequencies, shown as shaded areas in Fig. \ref{fig3}(b). Green and gold areas correspond respectively to higher and lower frequencies of the 1.15 MHz peak. The ghost image, calculated according to Eq. (\ref{eq:gi}) using the side windows does not show any structure, due to the lack of physical intensity correlation between the two beams. On the contrary by selecting the window around the alias of the storage ring frequency, the ghost image clearly displays an oblique shadow. 
\begin{figure}[b]
\begin{center}
\includegraphics[scale=0.55]{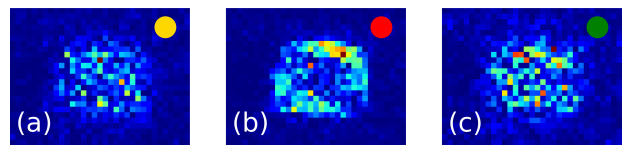}
\end{center}
\caption{Ghost imaging obtained by exchanging the role of reference and bucket. The transmitted beam is now the reference and the wire is moved to the diffracted beam, whose intensity distribution is integrated over an area of 20 $\times$ 30 pixels. The same Fourier filtering procedure described in Fig. \ref{fig3} is applied here and the results are displayed in panels (a-c). Panel (b) corresponds to selecting the window (red) centered on the alias frequency. The image size in all panels is 3.7 mm $\times$ 2.8 mm (H $\times$ V).} \label{fig4}
\end{figure}
That shadow is the ghost image of the wire, deformed according to the affine transform that relates the shape of the diffracted beam to the incident beam shape. 
\newline
As a further verification of the mechanism, we repeated the experiment after moving the wire in the diffracted beam, i.e. by exchanging the role of bucket and reference signal. The same data analysis procedure -- windowing either the alias or $f_{r/2}$ peak -- before ensemble averaging generates the ghost image shown in Fig. \ref{fig4}. Panels (a) and (c) represent the image obtained when windowing just below or above the alias frequency 1.15 MHz. Panel (b) is the image obtained when the window contains the alias frequency. \\
Two important differences between this image and the one in Fig. \ref{fig3}(d) are observable. The first difference originates from the position of the bucket beam. In this second case, the position of the bucket beam is variable due to the mechanical vibrations of the crystal discussed before. As a consequence, the sample is illuminated by a variable beam and the ghost image in Fig. \ref{fig4}(b) appears noisier than the corresponding image in Fig. \ref{fig3}(d). This problem is not present in the first case, as the position of the transmitted beam is not affected by mechanical vibrations of the crystal. \newline
The second difference is in the orientation of the ghost image. Having switched the role of bucket and reference beam, the affine transformation relating the ghost image to the actual sample image is inverted. Hence the apparent orientation of the wire in Fig. \ref{fig4}(b) is opposite to that in Fig. \ref{fig3}(d).
\newline
A final remark concerns the robustness of the GI mechanism studied here. In the ideal situation each speckle image acquired in a single frame must be generated with the x-rays produced from a single electron bunch. In practice however this is not strictly required. Some degree of mixing between the x-rays emitted by different pulses is acceptable, as long as the speckle visibility is not washed out. The sum of two (or few) speckle images is still a speckle image, and therefore ghost images can still be retrieved. For instance, as mentioned before, we successfully retrieved a ghost image by windowing the $f_{r/2}$ peak at 0.72 MHz, corresponding to considering the weighted average of two consecutive pulses. 
\newline
Regardless of the camera frame rate, light mixing between pulses occurs as a consequence of both the electronic noise in detection, and the scintillator's afterglow. X-rays are indirectly detected by scintillator screens \cite{rodnyi1997}. The process of scintillation has a characteristic decay time which is in general much longer than the duration of a single x-ray pulse; in fact it can be of the same order as the time separation between two consecutive pulses. The primary decay constants of the scintillation process for the scintillator we used were 630 ns and 70 ns for the CsI:Na and the YAG:Ce respectively. Therefore, even if the timing of the camera is adequate to select individual pulses, the image of a single pulse always contains residual intensity from the previous pulses, plus a constant background due to electronic noise. The background contribution however is filtered out during the Fourier processing, which eliminates all frequencies outside the selected window, and therefore does not contribute to the ghost imaging reconstruction procedure. The residual mixing between consecutive pulses is therefore purely limited to the speckle contribution. \\
In conclusion, we reported the experimental demonstration of direct ghost imaging using hard x-rays. The protocol was enabled by detecting the natural speckles present in the x-ray emission from a single electron bunch traveling in an undulator. A beam splitter was used to generate two copies of the beam, and both copies were simultaneously detected by a high speed camera. The camera frame rate was high enough to nearly isolate the x-rays from a single bunch or the average of two consecutive bunches. The sample (an x-ray opaque Cu wire) was placed in one of the beams, whose image was spatially integrated to constitute a point (bucket) detector. The ghost image of the wire was recovered under two configurations, using the intensity correlation between the bucket signal and the image of the empty beam. \\
The experimental demonstration of direct x-ray ghost imaging is extremely interesting for potential applications in medical imaging and ultrafast x-ray studies using free electron lasers. In both cases ghost imaging may represent an avenue to reduce radiation dose on the sample by using a suitably weak bucket beam, and maintaining the speckle correlation with the reference beam. 

The authors gratefully acknowledge M. Rutherford (Imperial College London) and Photron UK for kindly lending us the Photron camera used in this experiment. MS acknowledges the support of the French National Research Agency (ANR) via the EQUIPEX grant ANR-11-EQPX-0031.

\end{document}